\documentclass{article}  
\usepackage{bigsky2009}
\usepackage{graphicx}
\usepackage{url}
\usepackage{hyperref}
\usepackage{xspace}
\usepackage{amssymb}
\frompage{000} \topage{000}                                              %|
\def\rightmark{Jets areas and background subtraction}
\def\leftmark{G. Soyez}

\newcommand{\fastjet}{{\sf FastJet}}
\newcommand{\jd}{{\rm JD}}
\newcommand{\ja}{{\rm JA}}
 
\title{Jet areas as a tool for background subtraction} 
\authors{
{Gr\'egory Soyez$^{1,a}$%
}\\[2.812mm]
{\normalsize
\hspace*{-8pt}$^1$ Brookhaven National Laboratory, Building 510, NY 11973 Upton, USA
}}

\abstract{
In all modern hadronic colliders, jets recieve a large contribution
from a soft background: pileup in the case of proton-proton collisions
at the LHC, or the underlying event for heavy-ion collisions at RHIC
or the LHC. In these proceedings, we present a generic and simple
method, based on jet areas, to subtract the contribution of the soft
background from the jets. This allows for more precise kinematic
reconstruction of jets in dense environments.
}

\keyword{Jets, QCD} 
\PACS{12.38.Aw, 12.38.Bx}
 
\begin{document}
 
\maketitle

\section{Introduction}\label{sec:intro}

In all recent colliders, jets have been useful objects in numerous
analysis involving large-$p_t$ hadrons in the final state. However,
with the high-luminosity expected at the LHC as well as with the large
multiplicities obtained in heavy-ion collisions at RHIC, jets get
significant contributions from the soft background leading to biased
measures of their momentum. 

For all the analysis involving jets, it is important to determine as
well as possible their kinematic properties. therefore, one has to
develop techniques allowing to deal with the soft-background
contribution to the jet. 

In these proceedings, we discuss soft background subtraction using jet
areas. The underlying idea to this method is that, by suitably
defining what the area of a jet is, the contamination due to soft
background will be proportional to it.

We shall start by giving two possible definitions of the concept of
the area of a jet and show that those definitions allow for analytic
computations in perturbative QCD. We shall then explain how one can
use the jet area to subtract background contamination. 

Details concerning the definition of jet areas and their properties
can be found in \cite{areas}, and the application to background
subtraction in \cite{subtraction}.

\section{Jet areas}\label{sec:area}

\subsection{Definition}\label{sec:area_def}

Keeping in mind that we ultimately want to use jet areas in order to
correct from soft-background contamination, we aim at a definition
that mimics the reaction of the jet to soft particles. To do that in
practice, we will introduce additional, infinitely soft, particles
that we shall refer to as {\it ghosts}. For infrared-safe algorithm,
the addition of ghosts will not alter the clustering. We can then
define the area of a jet as the region in rapidity-azimuth where it
catches ghosts.

We can consider two different ways of doing this:
\begin{itemize}
\item {\bf Passive area}: the passive area is defined as the region in
  which a jet would catch a single ghost added to the event.
\item {\bf Active area}: in this case, we add to the event a set of
  ghosts $\{g_i\}$, with density per unit area $\nu_g$. If $n_J$
  ghosts are clustered with a jet $J$, its area w.r.t. to that set of
  ghosts will be $n_J/\nu_g$. We then define the active jet area of
  $J$ as the limit of $\langle n_J/\nu_g \rangle$, averaged over all
  ghosts distributions, when $\nu_g$ goes to infinity, {\em i.e.} in
  the limit of infinitely dense ghost coverage. Note that in this
  case, one can also end up with jets only made of ghosts.
\end{itemize}

Both the passive and the active area can easily be defined as
4-vectors. Those are formally defined by summing (or more precisely
integrating) the momenta of all the ghosts contributing to the area of
a jet, normalised in such a way that its transverse momentum
corresponds to the scalar area. 

Physically speaking, the passive area which clusters a single ghost
at a time has similar behaviours as a point-like background, while the
active areas correspond to diffuse, uniform, backgrounds.

A final comment concerns the jet algorithm choice. Since the definitions
of jet areas given in this Section involve adding soft particles, it
is of prime importance that these ghosts do not modify the clustering
of the hard particles in the original event, a property known as
infrared-safety. In what follows, we shall consider 4
infrared-and-collinear-safe algorithms: the $k_t$ \cite{kt},
Aachen/Cambridge \cite{cam} and anti-$k_t$ \cite{antikt} algorithms as
well as SISCone \cite{siscone}.

\subsection{Perturbative properties}\label{sec:area_props}

In this Section, we briefly highlight some nice perturbative
properties that can be explicitly computed for jet areas. The quantity
we shall be interested in is the average active and passive area of a
jet at the first non-trivial order in $\alpha_s$ {\it i.e.} including
the radiation of a gluon. We will work in the soft-collinear limit
that is relevant for small values of $R$ and properly take into
account the running of the QCD coupling constant.

The first step is to start with the area of a hard jet made of a
single particle of transverse momentum $p_{t1}$. The case of passive
areas is particularly simple: for every algorithm, an infinitely soft
particle will be clustered with the hard one if-and-only-if it is at
most at a distance $R$ from it, hence an area of $\pi R^2$.

The case of active areas is a bit more subtle. For the anti-$k_t$
algorithm, ghost recombinations with the hard particle will happen at
the very beginning of the cluster sequence, resulting in a area of
$\pi R^2$. For SISCone, one first has to notice that the stable cones
in the event made with a single hard particle and a dense ghost
coverage are (i) the cone centred on the hard particles and (ii) any
cone made only of ghosts. Through the split--merge phase, the
overlapping between the hard stable cone and all the pure-ghost ones,
with centres approaching the hard one up to a distance $R$, will lead
to a splitting (for $f\gtrsim 0.4$), leading to a cone of radius $R/2$
as the final jet, hence an area of $\pi R^2/4$. Finally, in the case
of the $k_t$ and Cambridge algorithms, ghosts will also cluster among
themselves leading to different areas for different sets of ghosts. On
average, one finds an area around $0.81 \pi R^2$ for both $k_t$ and
Aachen/Cambridge (with respective dispersions of $0.28 \pi R^2$ and
$0.26 \pi R^2$).

\begin{figure}
\includegraphics[angle=270,width=\textwidth]{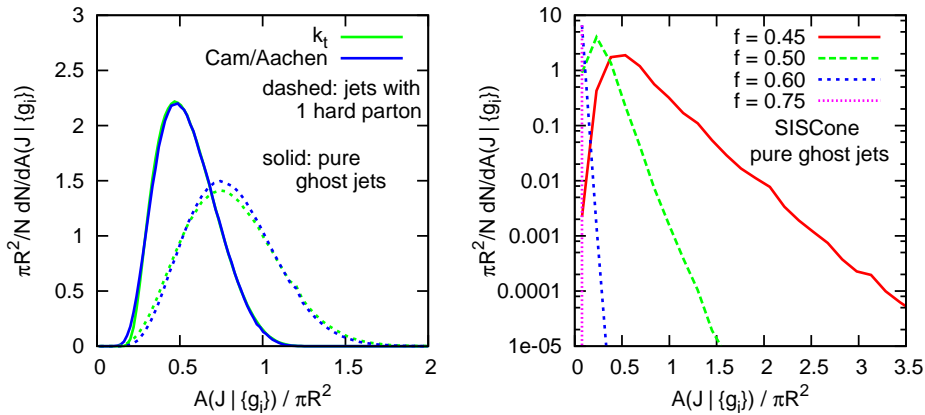}
\caption{Distribution of the active areas for a hard jet made of a
  single hard particle and the pure-ghost jets. Left: $k_t$ and
  Aachen/Cambridge, right: SISCone.}\label{fig:area_1part}
\end{figure}

The distribution areas for the hard jet and the pure-ghost jets are
presented in Figure \ref{fig:area_1part}. Note that for SISCone, for
small values of the overlap threshold there is a substantial
probability that pure-ghost jets become very large (known as {\it
  monster jets}). To avoid this, larger values of $f$ can be
chosen. This kind of arguments suggests $f=0.75$ as a sensitive default.

Next, we proceed with the computation of jet areas for situations with
a hard particle of transverse momentum $p_{t1}$ and a softer one with
transverse momentum $p_{t2}\ll p_{t1}$ located at a distance $\Delta$
from the first one. The calculation of the area as a function of
$\Delta$ goes typically as for the 1-particle case, though with
slightly more involved geometry, so we will not detail the results
here (see \cite{areas} for details). In the case of the passive area,
one can perform all the computations analytically, while for passive
areas, anti-$k_t$ and SISCone are treatable analytically but we once
again have to rely on numerical simulations to account for the
ghost-distribution dependence in the two remaining cases.

The important point is that, once we have the results for 1 and 2
particle jets, we can compute the average area of a jet perturbatively
at order $\alpha_s$. At leading order, a jet is made of a single
parton, while at next-to-leading order in $\alpha_s$, an extra gluon
can be radiated. For a jet definition $\jd$ one thus has
\[
\left\langle A_\jd \right \rangle \simeq \left\langle A_\jd \right
\rangle_{\rm 1 hard} + \int d\Delta \int_{Q_0/\Delta}^{p_{t1}} dp_{t2}
\frac{dP}{d\Delta\, dp_{t2}} \left(
\left\langle A_\jd(\Delta) \right \rangle_{\rm 2 part.}-
\left\langle A_\jd(0) \right\rangle_{\rm 2 part.}
\right),
\]
where the last term receives a 2-particle contribution from real-gluon
emissions and a 1-particle virtual correction. Since the $p_{t2}$
integration has a soft divergence --- coming from the gluon-radiation
probability --- one has to introduce a cut-off. The value $Q_0/\Delta$
above comes by requiring that the transverse momentum of the emitted
gluon relative to its parent, {\em i.e.} $p_{t2}\Delta$ in the
small-angle approximation, is larger that the soft cut-off $Q_0$. We
will come back to the importance of this cut-off later on.

In the soft and collinear approximation, the probability for gluon
emissions is
\[
\frac{dP}{d\Delta\, dp_{t2}}
 = C_R \frac{2\alpha_s(p_{t2}\Delta)}{\pi}
   \frac{1}{\Delta}\frac{1}{p_{t2}}
\]
with the colour factor $C_R$ is $C_F$ or $C_A$ for quark and gluon
jets respectively. Again, the argument of the coupling is the
transverse momentum of the second particle relative to the first one.

\begin{table}
\centerline{
\begin{tabular}{|l|c|c|c|c|}
\hline
 & \multicolumn{2}{c|}{$a_{\ja,{\rm 1 hard}}$} &
 \multicolumn{2}{c|}{$d_\ja$} \\
\cline{2-5}
           & passive & active & passive & active \\
\hline
$k_t$      &    1    &  0.81  &   0.56  &  0.52  \\
Cam/Aachen &    1    &  0.81  &   0.08  &  0.08  \\
anti-$k_t$ &    1    &   1    &   0     &  0 \\
SISCone    &    1    &  1/4   &  -0.06  &  0.12  \\
\hline
\end{tabular}
}
\caption{Summary table giving the one-particle average area and the
  scaling violation coefficients for the passive and active areas and
  for various jet algorithms. All quantities correspond to an area
  divided by $\pi R^2$.}\label{tab:areasummary}
\end{table}

The final result, normalised by $\pi R^2$, can be cast under the form
\begin{equation}\label{eq:scalingviol}
\frac{\left\langle A_\jd \right \rangle}{\pi R^2}
 \simeq a_{\ja,{\rm 1 hard}}
 + d_\ja \frac{C_R}{b_0\pi} \log\left( 
   \frac{\alpha_s(Q_0)}{\alpha_s(Rp_{t1})}
   \right),
\end{equation}
where $b_0=(11C_A-2n_f)/(12\pi)$ is the one-loop QCD beta-function,
and we have neglected terms that were not logarithmically enhanced. To
grasp the physical content of this equation, a few comments are in
order:
\begin{itemize}
\item The coefficients $a_{\ja,{\rm 1 hard}}$ and $d_\ja$, related to
  the one-particle area and the one-gluon emission corrections
  respectively, depend on the chosen algorithm $\ja$ and area
  type. They can both be analytically computed from the one and
  two-particle situations mentioned above. A summary of the relevant
  values is presented in Table \ref{tab:areasummary}.
\item Not only the area of a jet is not always $\pi R^2$ as one might
  naively expect --- we already noticed that earlier --- but
  eq. (\ref{eq:scalingviol}) shows that the jet substructure generates
  some scaling violations. 
\item Because of the prefactor $C_R$, the scaling violations will be
  larger for gluon jets than for quark jets.
\item The cut-off $Q_0$ is a non-perturbative scale. It is the trace
  that, since the addition of a soft particle can modify them, jet
  areas are not infrared-safe quantities. Physically, they should be
  regularised by a scale related to the underlying event density in
  $pp$ or heavy-ion collisions, or by the pileup density in the
  presence of pileup. It is interesting to notice that, as long as we
  keep the limit $p_{t1}\gg Q_0$, the scaling violations reduce when
  increasing the background density.
\item In the specific case of the anti-$k_t$ algorithm, the
  coefficient $d_{{\rm anti-}k_t}$ vanishes. Actually, in the limit of
  soft emissions, this is true at any order and the area remains $\pi
  R^2$, a fact reminiscent of the rigidity of the algorithm. The
  correction to that result will only come with power-suppressed
  factors, without logarithmic enhancement.
\end{itemize}

\subsection{Implementation}\label{sec:area_impl}

In practice, jet areas have been implemented in \fastjet
\cite{fastjet}. Ghosts are placed on a square grid with each node
slightly shifted. The binning of that grid, corresponding to the
quantum of area carried by each ghost, can be specified to achieve the
desired accuracy. The computation can also be averaged over different
ghosts distributions, {\it i.e.}  different shifts of the grid.

\section{Background subtraction}\label{sec:subtraction}

We finally come to the question of how jet areas can be used to
subtract jet contamination from soft background. For simplicity, we
shall concentrate on the case of a background uniform over a rapidity
range $|y|\le y_{\rm max}$. If the background has a density per unit
area $\rho$, the subtraction formula for a given jet is the following:
\begin{equation}\label{eq:subtraction}
p^\mu_{\rm subtracted} = p^\mu_{\rm jet}-\rho A^\mu_{\rm jet},
 \end{equation}
where $A^\mu$ is the jet 4-vector area. We discuss below the two
building blocks of this formula, namely the jet area and the
background density.

\subsection{The jet area}

This point has already been addressed in Section \ref{sec:area}. Given
the fact that the background is uniform, active areas are a natural
choice. However, for large multiplicities, the passive area tends to the
active one, so the passive area can also be used. This might be
relevant {\em e.g.} for SISCone as the computation of passive areas is
much less time-consuming than the one of active areas.

\subsection{The medium density per unit area}

\begin{figure}
\centerline{\includegraphics[width=0.5\textwidth]{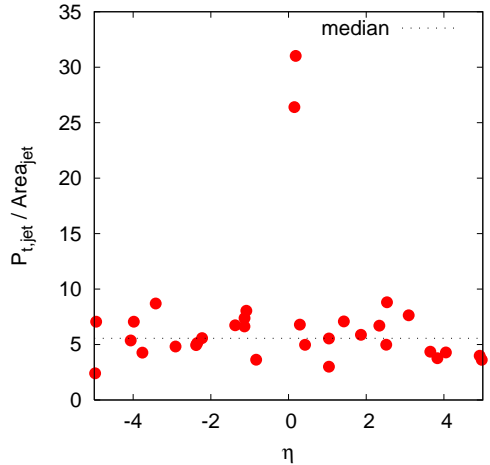}}
\caption{Example of the distribution of $p_t$/area of the jets for one
specific example. The lower points are pure-background jets while the
2 upper ones are the hard jets.}\label{fig:area_scatter}
\end{figure}

We are left with the estimation of the background density per unit
area. The estimation we propose here is based on the observation that
the ratio of the $p_t$ of a jet divided by its area can behave in two
distinctive ways: it will be around $\rho$ for the many jets made
purely of background, and much larger for the few hard jets. As a
consequence, the median of the set of $p_t$/area for all the jets in
the event gives an estimate of the background density $\rho$. This is
illustrated in Figure \ref{fig:area_scatter}.

A tricky point here is that we optimally want to avoid having too many
jets with small area as it would lead to large uncertainties in
$p_t$/area. In that respect, the $k_t$ and Cambridge/Aachen algorithms
are the best suited choices. This means that, whatever algorithm you
plan on using and apply the subtraction method, it is advised to
pre-estimate $\rho$ using $k_t$ or Cambridge/Aachen, and then use that
value of $\rho$ to perform the subtraction with the chosen algorithm.

\subsection{Properties and applications}

\begin{figure}
\includegraphics[width=0.5\textwidth]{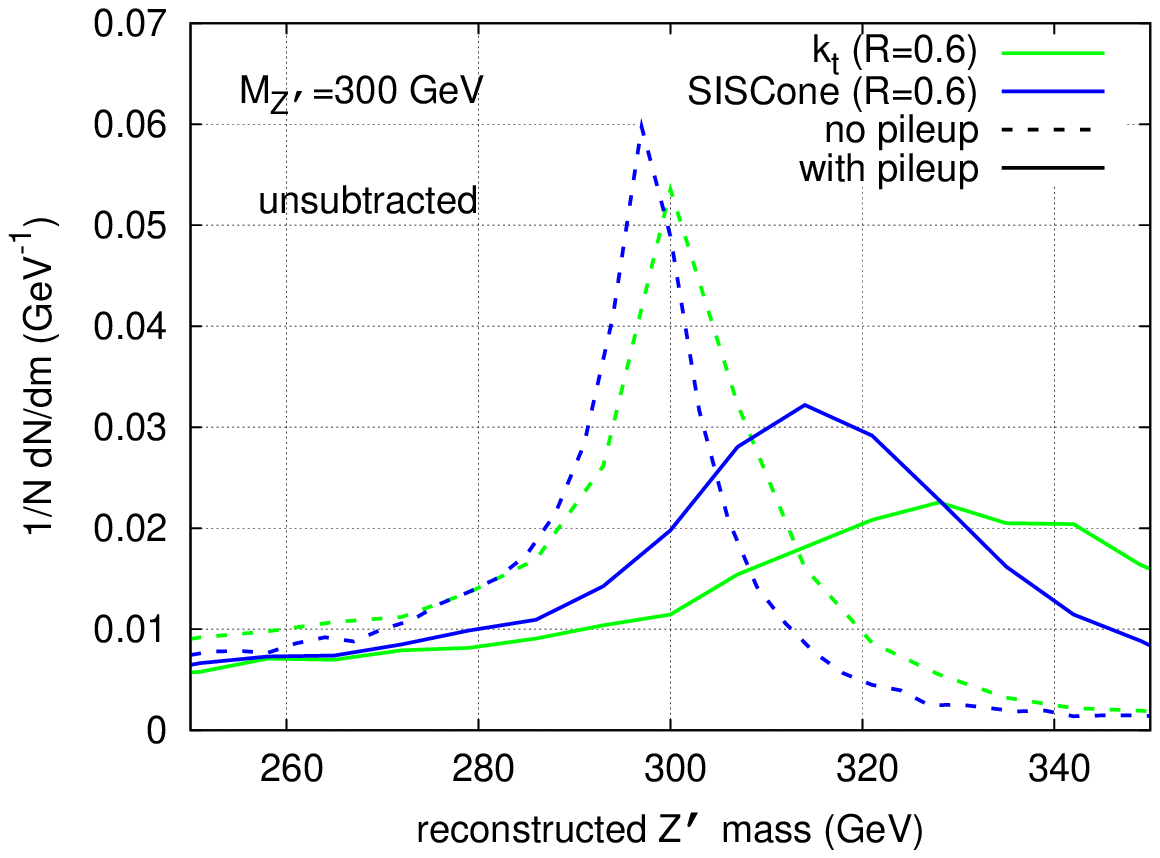}%
\includegraphics[width=0.5\textwidth]{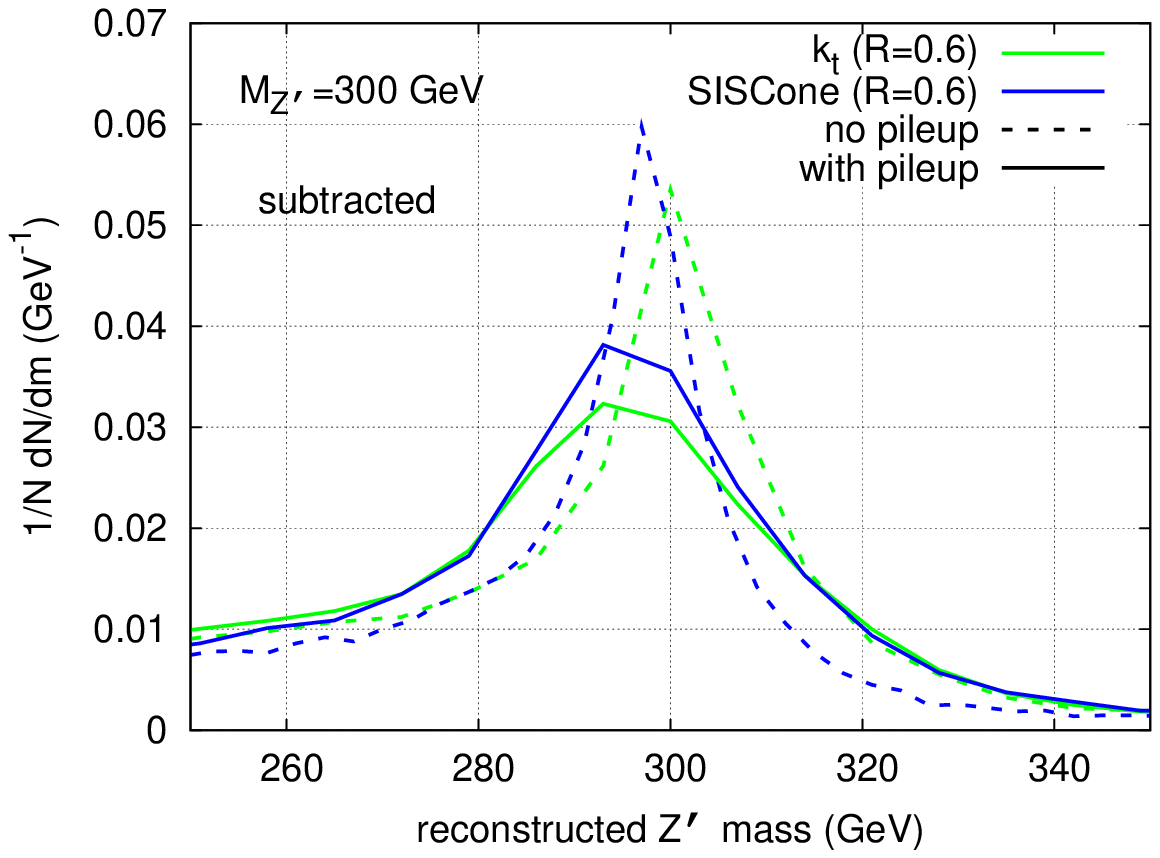}
\caption{Left: reconstruction of the $Z'$ mass peak without pileup
  (dashed lines) and with pileup without applying subtraction (solid
  lines). Right: same except that our subtraction method has been
  applied in the case where pileup was added. Both cases are presented
  for the $k_t$ and SISCone algorithms with a radius of
  0.6.}\label{fig:example}
\end{figure}

An essential point here is that this subtraction method can be applied
individually for each event. The main advantage is that it
significantly reduces the event-by-event fluctuations of the density
of the background, reducing its smearing effects.

In addition, the method is simple and general enough to have a broad
range of applications, ranging from pileup subtraction in $pp$
collisions to underlying-event subtraction in $pp$ or $AA$
collisions. See {\em e.g.} \cite{star} for an explicit experimental
application by STAR for $AA$ collisions at RHIC.

Finally, let us illustrate the effects of subtraction on a simple
example. We have generated with Pythia 6.4 (tune DWT), a set of
fictitious $Z'$ events, where the $Z'$ has a mass fixed to 300 GeV
(with a narrow width of less that 1 GeV) and decays into a $q\bar q$
pair$^{b}$. The events are clustered and the two
hardest of the resulting jets, the best candidates to match the
original quark and anti-quark, are used to reconstruct the $Z'$. We
can study how the mass peak of the $Z'$ is reproduced in 3 different
cases:
\begin{enumerate}
\item without pileup addition,
\item with pileup, simulated by adding minimum bias events with a
  Poisonian distribution corresponding to a luminosity of 0.25
  mb$^{-1}$ per bunch crossing (LHC at designed luminosity), 
\item with pileup and applying the subtraction method presented here.
\end{enumerate}

The results of the reconstruction of the $Z'$ mass peak are presented
in figure \ref{fig:example} without pileup subtraction on the left and
with pileup subtraction on the right. In both cases, we show the
results when the $k_t$ and SISCone algorithms with a radius of 0.6 are
used for the clustering. The subtraction on the right plot has been
done using~(\ref{eq:subtraction}) where $\rho$ is estimated using the
$k_t$ algorithm with a radius of 0.5, keeping all the jets with
$|y|<5$. While without subtraction the position of the peak is
severely shifted towards larger masses and its width is much larger,
after applying our subtraction technique, the peak comes back to a
good position and its width, though a bit larger than without pileup,
is much reduced compared to what we had before subtraction.

\section*{Acknowledgements}
I am very grateful to Matteo Cacciari and Gavin Salam for
collaboration on the topics presented in these proceedings.
 
\section*{Note(s)}
\begin{notes}
\item[$a$]
  Supported by Contract No. DE-AC02-98CH10886 with the U.S. Department of Energy.\\
  E-mail: gsoyez@quark.phy.bnl.gov \label{note:doe}
\item[$b$] Such a small mass for a potential $Z'$ has already been
  excluded. However, we use it here simply as a source of dijets at a
  fixed scale.\label{note:zprime}
\end{notes}

\vfill\eject
\end{document}